\documentclass[twocolumn,showpacs,amsmath,aps]{revtex4}
\usepackage{graphicx,color}
\usepackage{CJK}

\newcommand{\nl}{\nonumber \\}

\newcommand{\be}{\begin{equation}}
\newcommand{\ee}{\end{equation}}
\newcommand{\bea}{\begin{eqnarray}}
\newcommand{\eea}{\end{eqnarray}}

\newcommand{\Eq}[1]{Eq.\,(\ref{#1})}
\newcommand{\Eqs}[1]{Eqs.\,(\ref{#1})}

\newcommand{\la}{\langle}
\newcommand{\ra}{\rangle}
\newcommand{\dg}{\dagger}

\newcommand{\mb}{\mbox}
\newcommand{\ep}{\epsilon}

\newcommand{\al}{\alpha}

\begin{document}
\begin{CJK*}{GBK}{Song}

\title{ Large-deviation analysis for counting statistics
              in mesoscopic transports }

\author{Jun Li$^1$, Yu Liu$^1$, Jing Ping$^1$, Shu-Shen Li$^1$,
         Xin-Qi Li$^{2,1}$\footnote{E-mail: lixinqi@bnu.edu.cn},
         and YiJing Yan$^3$}
\affiliation{$^1$State Key Laboratory for Superlattices and
Microstructures, Institute of Semiconductors, Chinese Academy of
Sciences, P.O.~Box 912, Beijing 100083, China}
\affiliation{$^2$Department of Physics, Beijing Normal University,
  Beijing 100875, China}
\affiliation{$^3$Department of Chemistry, Hong Kong University of Science
and Technology, Kowloon, Hong Kong}

\date{\today}

\begin{abstract}
We present an efficient approach, based on a number-conditioned master
equation, for large-deviation analysis in mesoscopic transports.
Beyond the conventional full-counting-statistics study,
the large-deviation approach encodes complete information of both
the typical trajectories and the rare ones, in terms of revealing
a continuous change of the dynamical phase in trajectory space.
The approach is illustrated with two examples:
(i) transport through a single quantum dot, where we reveal
the inhomogeneous distribution of trajectories in general case
and find a particular scale invariance point in trajectory statistics;
and (ii) transport through a double dots, where we
find a dynamical phase transition between two distinct phases
induced by the Coulomb correlation and quantum interference.
\end{abstract}



\maketitle

\section{Introduction}

It has been widely recognized that, beyond the usual average current,
the current fluctuations in mesoscopic transport can provide as well
very useful information for the nature of transport mechanisms \cite{rev0003}.
In particular, a fascinating theoretical approach, known as full counting
statistics (FCS)\cite{Lev9396}, can conveniently yield
all the statistical cumulants of the number of transferred charges
through a variety of mesoscopic systems
\cite{Naz01,But02,Sam05,Sch0506,Bel05,Bk00,Bel04,Bla04,Jau04,Bee06,Li07}.
Experimentally, the real-time counting statistics has been realized
in transport through quantum dots \cite{Ens06},
representing a crucial achievement of being able to count individual
tunneling events in mesoscopic transport.

For mesoscopic transport, in addition to the well-known Landauer-B\"uttiker
scattering theory \cite{Dat95} and the nonequilibrium Green's function
formalism \cite{Jau96},
a particle-number-resolved master equation approach
was demonstrated very useful in studying quantum noise in transport
\cite{Gur96,Sch01,Li05,Moz0204,Gur03,Jau05}.
The key quantity in this approach is the particle-number($n$)-dependent
density matrix (i.e., $\rho^{(n)}(t)$,
the reduced state of the central system in the transport setup).
The connection of $\rho^{(n)}(t)$ with the distribution function
in counting statistics simply reads $P(n,t)=\mb{Tr}[\rho^{(n)}(t)]$,
where the trace is over the central system states.
From this distribution function, all orders of cumulants
of the transmission electrons can be calculated,
most conveniently, by introducing the cumulant generating function (CGF)
$ e^{-\tilde{\cal F}(\chi,t)}=\sum_n P(n,t)e^{i n\chi}$.
This is in fact a discrete Fourier transformation,
and $\chi$ corresponds to the so-called {\it counting field}.

Instead of the above discrete Fourier transformation,
we may consider a replacement: $e^{i\chi n}\Rightarrow e^{-xn}$.
That is, we introduce the {\it dual}-function of $P(n,t)$:
$ P(x,t)=\sum_n e^{-x n} P(n,t)=e^{-{\cal F}(x,t)} $.
The real nature of the transforming factor $e^{-x n}$,
in contrast to the complex one, $e^{i\chi n}$,
makes the resultant $P(x,t)$ in some sense resemble
the partition function in statistical mechanics.
This function, in broader contexts, is named
{\it large-deviation} (LD) function,
which, remarkably, allows to access the {\it rare fluctuations}
or {\it extreme events} in random and dynamical systems
when performing the LD analysis \cite{LD9809}.
The partition function, in equilibrium statistical mechanics \cite{SM8792},
measures the number of microscopic configurations accessible to the system
under given conditions.
The LD function, analogously, is a measure of the number of trajectories
accessible for the dynamical aspects of a nonequilibrium system
\cite{LD9809}.
Relating to mesoscopic transports, if we are interested in the
{\it dynamical aspects} of the transport electrons in {\it time domain},
again, analogous to the partition function, the LD function is a measure
of the number of trajectories accessible to the electron ``counter".
Using the terminology in LD analysis, the trajectories in mesoscopic transports
are categorized by the dynamical order parameter ``$n$" or its conjugate
field ``$x$". Trajectories are given an exponential weight, $e^{-x n}$,
playing a role similar to the Boltzmann factor in statistical mechanics.

Since the LD theory describes the trajectory space
from multiple angles through the effect of the conjugate fields,
say, analyzing the ``$x$" dependence instead of limiting
$\chi\rightarrow 0$ as in the FCS study,
the LD approach is expected to be more instrumental
in studying the statistics of quantum systems.
In a most recent work \cite{Gar10}, the LD analysis was applied
to a number of well-known simple quantum optical systems.
Unexpectedly, striking dynamical phenomena such as scale invariance
of trajectories as well as dynamical phase transition were uncovered
for the statistics of photon emissions.
These findings indicate that more surprising and complex dynamical
behaviors are seemingly waiting to be discovered in various scenarios.
In the present work, we formulate a convenient LD method
for mesoscopic transports and apply it to a couple of examples.

The paper is organized as follows. In Sec.\ II we begin with a brief outline
of the particle-number ($n$)-conditioned master equation
together with a summary of the conventional FCS formalism based on it,
before formulating the LD approach to quantum transport.
As a first illustrative example, we present in Sec.\ III an LD analysis
for the simplest system of transport through a single quantum dot
where we reveal an inhomogeneous distribution of trajectories in general
and find a particular scale invariance point in trajectory statistics.
We then carry out in Sec.\ IV the second example of transport through
a parallel double dots where we find a dynamic phase transition between two
distinct phases induced by the Coulomb correlation and quantum interference.
Finally, in Sec.\ V we summarize the work.

\section{Large-Deviation Method}

\subsection{ Transport Master Equation }

In general, a mesoscopic transport setup can be described
by the following Hamiltonian
\bea\label{H-ms}
H &=& H_S(a_{\mu}^{\dg},a_{\mu})+\sum_{\al=L,R}\sum_{ k}
       \ep_{\al k}b^{\dg}_{\al k}b_{\al k}  \nl
  & &  + \sum_{\al=L,R}\sum_{\mu k}(t_{\al\mu k}a^{\dg}_{\mu}
       b_{\al k}+\rm{H.c.}) .
\eea
Here, $H_S$ is the Hamiltonian of the central system in between the electrodes,
with $a^{\dg}_{\mu}$ ($a_{\mu}$) the creation (annihilation) operator
of the (single particle) state ``$\mu$".
The second and third terms describe, respectively, the two (left and right)
electrodes and the tunneling between them and the central system.

Introducing
$F_{\mu} \equiv \sum_{\al k} t_{\al\mu k}b_{\al k}
  \equiv f_{L\mu} + f_{R\mu}$,
we re-express the tunneling Hamiltonian as
$H' = \sum_{\mu} ( a^{\dg}_{\mu} F_{\mu}
       + \rm{H.c.} ) $ .
Regarding the tunneling Hamiltonian as a perturbation,
and properly clarifying the {\it subspace} of electrode states
in association with the electron number transmitted through
the system, say, the electron number arrived to the right electrode,
we can obtain a number-conditioned master equation \cite{Li05},
which formally reads
\begin{align}\label{nME}
\dot{\rho}^{(n)}=&-i\mathcal{L}\rho^{(n)}
-\sum_{j=0,\pm 1} \mathcal{R}_j \rho^{(n+j)} .
\end{align}
Here, $\rho^{(n)}$ is the reduced density matrix of the {\it system}
conditioned on the electron number ``$n$" arrived to the right electrode.
The Liouvillian $\mathcal{L}$ is the well-known commutator
defined by the system Hamiltonian $H_S$.
The superoperators $\mathcal{R}_j$ describe electron tunneling
between the system and the electrodes,
with an explicit form referred to Ref.\ \cite{Li05}.
For the convenience of latter quotation, here we only mention an
expression for the tunneling rates, based on the Fermi's golden rule,
$\Gamma_{\alpha\mu}=2\pi \Omega_{\alpha}|t_{\alpha\mu}|^2$,
where $\Omega_{\alpha}$ are the density of states of the electrodes,
and $t_{\alpha\mu}$ the tunneling amplitudes under an approximation
of energy independence.

{\it Remarks.}---
{\it (i)}
The above master equation is up to the
second-order cummulant expansion for the tunneling Hamiltonian \cite{Yan98}.
While this approximation applies well to most dissipative systems
(e.g., in quantum optics), in quantum transport it actually limits the
applicability to the so-called {\it sequential tunneling} regime.
Generalization to higher order expansion of $H'$
is possible, but quite complicated \cite{Yan0508}.
{\it (ii)}
The connection of the particle-number-conditioned density matrix with
the distribution function in counting statistics is straightforward,
i.e., $P(n,t)=\mb{Tr}[\rho^{(n)}(t)]$,
where the trace is over the central system states.
We will see in next subsection that all the cumulants of current
can be calculated conveniently from this distribution function.

\subsection{ Counting Statistics }

With the knowledge of $P(n,t)$ from the $n$-dependent density
matrix $\rho^{(n)}(t)$, the current cumulant generating function (CGF)
can be constructed through
\begin{equation}\label{CGF}
e^{-\tilde{\cal F}(\chi,t)}=\sum_n P(n,t)e^{in\chi} ,
\end{equation}
where $\chi$ is the so-called counting field.
Based on the CGF, $\tilde{\cal F}(\chi,t)$,
the $k_{\rm th}$ cumulant can be readily carried out via
$C_k=-(-i\partial_{\chi})^k \tilde{\cal F}(\chi,t)|_{\chi=0}$.
As a result, one can easily check:
the first two cumulants, $C_1=\bar{n}$ and $C_2=\overline{n^2}-\bar{n}^2$,
give rise to the mean value and the variance of the transmitted electrons,
while the third cumulant (skewness), $C_3=\overline{(n-\bar{n})^3}$,
characterizes the asymmetry of the distribution function.
Here, $\overline{(\cdots)}=\sum_n (\cdots)P(n,t)$.
Moreover, one can relate the cumulants to measurable transport quantities,
e.g., the average current by $I =eC_{1}/t$,
and the zero-frequency shot noise by $S =2e^{2}C_{2}/t$.
Also, the important Fano factor is simply given by $F=C_{2}/C_{1}$,
which characterizes the extent of current fluctuations:
$F>1$ indicates a super-Poisson fluctuating behavior,
while $F<1$ a sub-Poisson process.

\subsection{ Large-Deviation Formalism}

Instead of the discrete Fourier transformation, \Eq{CGF},
which is employed in the conventional counting statistics,
let us now consider a replacement: $e^{i\chi n}\Rightarrow e^{-xn}$.
That is, we introduce the following {\it dual}-function of $P(n,t)$:
\bea\label{LD-1}
P(x,t)=\sum_n e^{-x n} P(n,t)=e^{-{\cal F}(x,t)}.
\eea
The real nature of the transforming factor $e^{-x n}$,
in contrast to the complex one, $e^{i\chi n}$,
makes the resultant $P(x,t)$ in some sense resemble the
partition function in statistical mechanics.
Using analogous language in statistical mechanics,
in \Eq{LD-1}, the trajectories are categorized
by a dynamical order parameter ``$n$" or its conjugate field ``$x$".
This is realized by the exponential weight similar to the Boltzmann factor,
with the dynamical order parameter playing the role of energy or magnetization
and the conjugate field the role of temperature or magnetic field.

The function $P(x,t)$, in broader contexts, coincides with
the {\it large-deviation} (LD) function which plays
an essential role in large-deviation analysis.
In statistical mechanics, the partition function measures the number of
microscopic configurations accessible to the system under given conditions.
For the mesoscopic transport under consideration, if we are interested
in the dynamical aspects of the transport electrons, the above insight
can lead to a LD analysis in {\it time domain}.
That is, the LD function is a measure of the number of trajectories accessible
to the ``counter", which favorably characterizes the trajectory space
from multiple angles through the effect of the conjugate field.
In particular, it allows to inspect the {\it rare fluctuations}
or {\it extreme events} by tuning the conjugate field ``$x$".

We emphasize that, if one performs only the conventional FCS analysis,
using either a complex transform factor $e^{i\chi n}$ or a real one $e^{-x n}$
would make no difference since at the end the limit
$ \chi (x) \rightarrow 0$ will be involved.
However, for the LD study, we must use the real factor $e^{-x n}$,
which has a role of categorizing (selecting) the trajectories.
It is right this type of selection which enables us to perform statistical
analysis for the fluctuations of {\it sub-ensembles} of trajectories.
For instance, $x>0$ implies mainly selecting the {\it inactive} trajectories
(with small $n$),
while $x<0$ prefers the {\it active} trajectories (with large $n$).
In particular, by varying $x$, the $x$-dependent statistics can
reveal interesting dynamical behaviors in {\it time-domain}.
In other words, based on the distribution function $P(n,t)$
which contains complete information of all the trajectories,
the LD approach, beyond the conventional FCS, captures more
information from $P(n,t)$ via the $x$-dependent cumulants.

Let us now turn to the technical aspect of the LD approach,
based on \Eq{nME}. Similar to transforming $P(n,t)$ to $P(x,t)$,
we introduce $\rho(x,t)=\sum_n e^{-xn}\rho^{(n)}(t)$.
Then, from \Eq{nME} we obtain
\bea\label{rho-xt}
\dot{\rho}(x,t) = \left[ -i{\cal L} - {\cal R}_0
  - e^{x}{\cal R}_{1} - e^{-x}{\cal R}_{-1} \right] \rho(x,t) .
\eea
This equation allows us to carry out the LD function $P(x,t)$,
via $P(x,t)={\rm Tr}[\rho(x,t)]$,
where the trace is over the central system states.
Accordingly, we obtain the generating function
${\cal F}(x,t)=-\ln P(x,t)$, for {\it arbitrary} counting time $t$.
Further, we can prove:
\begin{subequations}\label{F-k}
\begin{align}
{\cal F}_1(x,t)&\equiv \partial_x {\cal F}(x,t)
=\frac{1}{P(x,t)}\sum_n n e^{-xn} P(n,t)
  \equiv \la n \ra_x ,   \\
{\cal F}_2(x,t)&\equiv \partial^2_x {\cal F}(x,t)
=-\la (n-\bar{n}_x )^2\ra_x,
\end{align}
\end{subequations}
and more generally,
${\cal F}_k(x,t)\equiv \partial^k_x {\cal F}(x,t)
=(-)^{(k+1)}\la (n-\bar{n}_x )^k\ra_x$.
Here, for brevity, we used also the notation $\bar{n}_x$ for $\la n \ra_x$.
Using these cumulants, we can define a finite-counting-time
average current $I(x,t)=e{\cal F}_1(x,t)/t$, and the
shot noise $S(x,t)=2e^2 {\cal F}_2(x,t)/t$.
Also, the generalized Fano factor, $F(x,t)={\cal F}_2(x,t)/{\cal F}_1(x,t)$,
will be of interest to characterize the fluctuation properties.

We notice that, in order to obtain ${\cal F}_k(x,t)$,
we need only to carry out the various $k$-th order derivatives of $P(x,t)$,
$P_k(x,t)=\partial^k_x P(x,t)$.
This observation allows us an efficient method to
compute the $x$-dependent cumulants for finite counting time.
That is, performing the derivatives $\partial^k_x$ on \Eq{rho-xt}
and defining $\rho_k(x,t)=\partial^k_x \rho(x,t)$,
we get a set of coupled equations for $\rho_k(x,t)$,
whose solution then gives $P_k(x,t)={\rm Tr}[\rho_k(x,t)]$.

In long counting time limit, it can be proved that
${\cal F}(x,t)\simeq t\lambda(x)$.
{\it In the remaining parts of this work, we will also call
$\lambda(x)$ as an LD function, with no more distinction from $P(x,t)$.}
Desirably, we notice that the asymptotic form, $P(x,t)\simeq e^{-t\lambda(x)}$,
allows for a simpler way to get the LD function $\lambda(x)$.
That is, one can identify it to the {\it smallest} eigenvalue
of the counting matrix, given by the r.h.s of \Eq{rho-xt}.

In this context, we would like to mention that the limit of
$x\rightarrow 0$ simply gives ${\cal F}_k(x,t)|_{x\rightarrow 0}=C_k$,
recovering the usual FCS cumulants.
Again, we remark that the LD function around $x=0$ encodes
information of the {\it typical} trajectories,
while away from $x=0$, on the other hand,
it encodes information about the {\it rare} trajectories.
For nonzero $x$, the possible anomalous behavior of $\lambda(x)$ with $x$,
as we will see in the following examples,
is usually a signature for nontrivial set of dynamical trajectories.
For instance, singularities of $\lambda(x)$ correspond to certain
dynamical (or space-time) phase transition, i.e., a rapid crossover
between two distinct dynamical phases.



\section{Transport through a Single-Level Quantum Dot}

As a preliminary example, let us consider the transport through
a single-level quantum dot \cite{Ens06}.
Also, we assume spinless electron for simplicity but not losing
physics in the sequential tunneling regime (large bias limit).
In this case the system Hamiltonian in \Eq{H-ms}
has a simple form $H_S=E_0 a^{\dg}a$,
which implies only two states involved
in the transport process, i.e.,
the empty state $|0\ra$ and the occupied one $|1\ra$.
Using this state basis, the $n$-dependent rate equation reads \cite{Li05}
\bea\label{rho-0011}
\dot{\rho}^{(n)}_{00}&=&-\Gamma_L\rho^{(n)}_{00}+\Gamma_R\rho^{(n-1)}_{11}, \nl
\dot{\rho}^{(n)}_{11}&=&-\Gamma_R\rho^{(n)}_{11}+\Gamma_L\rho^{(n)}_{00}, \nl
\dot{\rho}^{(n)}_{01}&=&-\frac{1}{2}(\Gamma_L+\Gamma_R)\rho^{(n)}_{01} ,
\eea
where $\Gamma_{L,R}$ are the well known tunneling rates given by
Ferm's golden-rule.
Now, transforming \Eq{rho-0011} in terms of
$\rho(x,t)=\sum_{n}\rho^{(n)}(t)e^{-x n}$
and transposing $\rho(x,t)$ into a column-vector form
$[\rho_{00},\rho_{11},\rho_{01}]^{T}$,
we formally obtain
\bea
 \dot{\rho}(x,t)= -\mathcal{M}(x) \rho(x,t),
\eea
where $\mathcal{M}(x)$ is a $3\times 3$ matrix, explicitly,
with the following form
\begin{equation}\label{x-ME-SD}
\mathcal{M}(x) =
\begin{pmatrix}
\Gamma_{L} & -e^{-x}\Gamma_{R} & 0  \\
-\Gamma_{L} & \Gamma_{R}  & 0  \\
0 & 0&  \frac{\Gamma_{L}+\Gamma_{R}}{2}
\end{pmatrix}  .
\end{equation}
For the purpose of an analytic analysis, we carry out
the eigenvalues of $\mathcal{M}(x)$.
They are, respectively,
$\lambda_{1}= (\Gamma_{L}+\Gamma_{R})/2$,
and $\lambda_{2,3}= (\Gamma_{L}+\Gamma_{R}\pm X)/2$,
where $X= \sqrt{ (\Gamma_{L}-\Gamma_{R})^2
+ 4 e^{-x}\Gamma_{L}\Gamma_{R} }$.
Since $X>0$, we can then identify that $\lambda_{3}$
is always the smallest eigenvalue for arbitrary ``$x$",
implying that the LD function $\lambda(x)=\lambda_{3}(x)$.

Knowing $\lambda(x)$, it is straightforward to calculate
the $x$-dependent current and Fano factor.
In Fig.\ 1 we plot $\lambda(x)$, $I(x)$ and $F(x)$
for the case of $\Gamma_{L}=\Gamma_{R}\equiv\Gamma$.
We find that the $x$-dependent current decreases with $x$.
This is because, through modifying the distribution function $P(n)$
to $P(n)e^{-nx}$, a larger $x$ defines a {\it less active} subensemble
of trajectories, which then gives a smaller {\it subensemble} average current.
However, very interestingly, for this particular (equal coupling)
configuration, we find in Fig.\ 1 an $x$-{\it independent} Fano factor,
which indicates that all subensembles of trajectories,
no matter how active or inactive, have the same fluctuation properties
of the typical trajectories.
This remarkable {\it scale invariance} of dynamical trajectories
is very unusual, holding only for $\Gamma_L=\Gamma_R$.

Actually, we can carry out the analytic result
of Fano factor for arbitrary coupling symmetry, which reads
$F(x) = 1-\frac{1}{2}(1+R e^x)^{-1}$,
where $R=(\Gamma_L-\Gamma_R)^2/4\Gamma_L\Gamma_R$.
Then, we see that, with the increase of $x$, the Fano factor would
increase and asymptotically reach the Poissonian limit of unity.
This means that, differing from $\Gamma_L=\Gamma_R$,
in the asymmetric setup the fluctuations will become stronger
when we sample the trajectories from active to inactive ones.
Finally, we notice that for highly asymmetric coupling,
e.g., $\alpha\equiv\Gamma_L/\Gamma_R >> 1$,
$F(x) \simeq 1-\frac{1}{2}(1+\frac{\alpha}{4} e^x)^{-1}\simeq 1$.
This corresponds to a single barrier tunneling limit,
which gives a Poissonian statistics for the tunneling events.

\begin{figure}[h]
\begin{center}
\includegraphics[width=6cm]{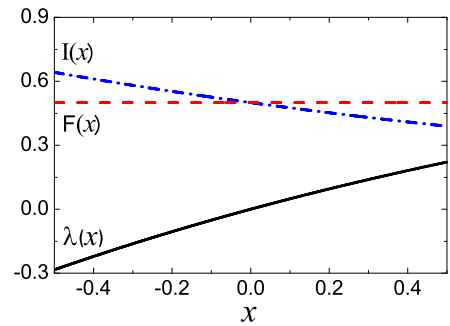}
\caption{\label{fig1}
(color online)
The LD function $\lambda(x)$, $x$-dependent current $I(x)$ and
Fano factor $F(x)$ for transport through a single-level quantum dot.
In the symmetric case of $\Gamma_{L}=\Gamma_{R}$ as assumed here,
the $x$-{\it independent} Fano factor indicates a remarkable
{\it scale-invariance} fluctuations of the dynamical trajectories. }
\end{center}
\end{figure}

\section{transport through double dots \
         in parallel }

\subsection{Model in a Transformed Representation}

In this section we consider a bit more complex setup which
consists of a double (quantum) dots (DD) connected in parallel
to two leads. Moreover, the interference loop is
pierced by a magnetic flux ($\Phi$), similar to
the well known Aharonov-Bohm interferometer.
It was shown recently that this system has some interesting properties
such as a non-analytic current switching behavior,
anomalous phase shift, and giant fluctuations of current
--- all of them induced by the interplay of inter-dot Coulomb correlation
and quantum interference \cite{LG09}.

Following Ref.\ \cite{LG09}, we assume that each dot has only
one level, $E_{1(2)}$, involved in the transport.
In large bias limit, for simplicity, we also neglect
the spin degrees of freedom, whose effect,
under strong Coulomb blockade,
can be easily restored by doubling the tunneling
rates of each dot with the left lead \cite{LG09}.
The entire system Hamiltonian reads
\bea\label{a1}
H=H_B+H_T+\sum_{\mu =1,2} E_\mu a_\mu^\dagger a_\mu
+Ua_1^\dagger a_1 a_2^\dagger a_2\, .
\eea
$a_{1,2}^\dagger$ are the creation operators for the DD,
and the last term describes the interdot repulsion.
The first term, $H_B=\sum_k [\epsilon_{Lk}b_{Lk}^\dagger b_{Lk}
+\epsilon_{Rk}b_{Rk}^\dagger b_{Rk}]$,
is for the leads and $H_T$ for their couplings to the dots,
\bea\label{a2}
 H_T=\sum_{\mu,k}\Big (t_{\mu L}a_\mu^\dagger b_{Lk}
 +t_{\mu R}b_{Rk}^\dagger a_\mu\Big )+{\rm H.c.}\, .
\eea
Here, we assume that the couplings of the dots to the leads,
$t_{\mu L(R)}$, are independent of energy.
In the absence of a magnetic field one can always choose the gauge
in such a way that all couplings are real.
In the presence of a magnetic flux $\Phi$, however, the tunneling
amplitudes between the dots and the leads are in general complex.
We write $t_{\mu L(R)}={\bar t}_{\mu L(R)}e^{i\phi_{\mu L(R)}}$,
where $\bar t_{\mu L(R)}$ is the coupling without the magnetic field.
The phases are constrained to satisfy
$\phi_{1L}+\phi_{1R}-\phi_{2L}-\phi_{2R}=\phi$,
where $\phi\equiv 2\pi\Phi/\Phi_0$.
In this work, we only consider the case with strong inter-dot
Coulomb blockade($E_{1,2}+U\gg \mu_{L}$)(two dots cannot be both
occupied simultaneously).

We found, in Refs.\ \cite{LG09}, that a transformed
basis can benefit the analysis a lot.
For simplicity, we consider $E_{1}=E_{2}$.
In this case, the SU(2) basis transformation reads
\bea
\left( \begin{array}{c}  \widetilde {a}_{1}\\
\widetilde {a}_{2}  \end{array} \right)
= {1\over \bar{t}_R} \left(\begin{array} {cc}t_{1R}
& t_{2R}\\
-t^*_{2R}&t^*_{1R} \end{array}\right)
\left (\begin{array}{c}a_{1}\\
a_{2}\end{array}\right) ,  \label{a11}
\eea
where $\bar{t}_R\equiv (\bar t_{1R}^2+\bar t_{2R}^2)^{1/2}$.
Under this transformation, the DD Hamiltonian,
$H_{\rm DD}=\sum_{\mu =1,2} E_\mu a_\mu^\dagger a_\mu
 + Ua_1^\dagger a_1 a_2^\dagger a_2$, is invariant,
while the tunneling Hamiltonian is transformed to
\bea
 H_T=\sum_{\mu,k}\Big (\widetilde{t}_{\mu L}\widetilde{a}_\mu^\dagger b_{kL}
 +\widetilde{t}_{\mu R}b_{kR}^\dagger \widetilde{a}_\mu\Big )
 +{\rm H.c.} ~, \label{bianhuan}
\eea
where, in the transformed basis, the tunneling amplitudes read
\bea
\widetilde t_{1R} &=& \bar{t}_R,
 ~~~~~~~~~~~~~~~~ \widetilde t_{2R}=0,  \nl
\widetilde t_{1L} &=& e^{i(\phi_{2L}+\phi_{2R})}(\bar t_{1L}
\bar t_{1R}\, e^{i\phi}+\bar t_{2L}\bar t_{2R})/\bar{t}_R,   \nl
\widetilde t_{2L} &=& -e^{i(\phi_{2L}-\phi_{1R})}(\bar t_{1L}
\bar t_{2R}\, e^{i\phi}-\bar t_{2L}\bar t_{1R})/\bar{t}_R .
\eea
The advantage of using the transformed basis is thus clear:
the state $\widetilde{a}^{\dagger}_2|0\ra$
becomes decoupled with the right lead, and its coupling with the left
lead is magnetic-flux tunable (in particular, can be switched off).

\subsection{Large-Deviation Function}

In the transformed basis, the strong Coulomb repulsion among
the inner dot and inter dots
defines a reduced Hilbert space expanded by
$\{ |0\rangle, |1\rangle, |2\rangle \}$,
where $|0\rangle$ stands for the vacant DD state, and $|1(2)\rangle$
for the single occupation state $\widetilde{a}^{\dagger}_{1(2)}|0\ra$.
Using this basis, the number($n$)-dependent master equation reads
\begin{align}\label{aa13}
\dot{\rho}^{(n)}_{00}=&-2\Gamma_{L}\rho^{(n)}_{00}
+2\Gamma_{R}\rho^{(n-1)}_{11} ,\nl
\dot{\rho}^{(n)}_{11}=&(1+\cos\phi)\Gamma_{L}\rho^{(n)}_{00}
-2\Gamma_{R}\rho^{(n)}_{11} ,\nl
\dot{\rho}^{(n)}_{22}=&(1-\cos\phi)\Gamma_{L}\rho^{(n)}_{00} ,\nl
\dot{\rho}^{(n)}_{12}=&i\sin\phi\Gamma_{L}\rho^{(n)}_{00}
-\Gamma_{R}\rho^{(n)}_{12}  , \nl
\dot{\rho}^{(n)}_{21}=&-i\sin\phi\Gamma_{L}\rho^{(n)}_{00}
-\Gamma_{R}\rho^{(n)}_{21} .
\end{align}
Here, again, ``$n$" is the electron number counted at the right lead.
Also, for the sake of brevity, we assumed that
$\bar {t}_{1 L}=\bar {t}_{2 L}=\bar {t}_{L}$,
and $\bar {t}_{1 R}=\bar {t}_{2 R}=\bar t_{R}$.
Then, the tunneling rates are defined, respectively,
as $\Gamma_{L}=2\pi \Omega \bar {t}_{L}^{2}$
and $\Gamma_{R}=2\pi \Omega \bar{t}_{R}^{2}$,
with $\Omega$ the density of states of the leads.

Similar to the single dot studied in the preceding section,
making the LD transformation,
$\rho(x,t)=\sum_{n}\rho^{(n)}(t)e^{-x n}$,
and transposing $\rho(x,t)$ into a column-vector form
$[\rho_{00},\rho_{11},\rho_{22},\rho_{12},\rho_{21}]^{T}$,
we can reexpress \Eqs{aa13} as
$ \dot{\rho}(x,t)= -\mathcal{M}(x) \rho(x,t)$,
with $\mathcal{M}(x)$ a $5\times 5$ matrix:
\begin{equation}\label{k-ME-2}
\mathcal{M}(x) =
\begin{pmatrix}
2\Gamma_{L} & -2\Gamma_{R}e^{-x}& 0 & 0 & 0\\
-(1+\cos\phi)\Gamma_{L} & 2\Gamma_{R}& 0& 0 &0\\
-(1-\cos\phi)\Gamma_{L}& 0& 0& 0& 0\\
-i\sin\phi ~\Gamma_{L}& 0& 0& \Gamma_{R}&0\\
i\sin\phi ~\Gamma_{L}& 0& 0& 0& \Gamma_{R}
\end{pmatrix} .
\end{equation}
Its eigenvalues read, respectively,
$\lambda_{1}=0$,
$\lambda_{2(3)}=\Gamma_{R}$,
and $\lambda_{4(5)}=(\Gamma_{L}+\Gamma_{R}) \pm X $,
where $X= \sqrt{ (\Gamma_{L}-\Gamma_{R})^2
+ 2e^{-x}\Gamma_{L}\Gamma_{R}(1+\cos\phi) }$.
Since $\lambda_{2,3,4}$ are positive, we can conclude that the {\it smallest}
eigenvalue is the smaller one among $\lambda_1$ and $\lambda_5$.
The analysis in Sec.\ IV (D) will show that, by varying the conjugate
field $x$, $\lambda_5$ has a crossover from negative to positive,
implying that $\lambda_1=0$ will suddenly become the smallest eigenvalue.
This behavior actually indicates a dynamic phase transition.
Before proceeding to such an analysis, for the sake of completeness,
in the following we first outline a scheme to account for
the dephasing effect between the dots.

\subsection{Dephasing Effect}

In real systems, dephasing may originate from surrounding environments,
which cause random fluctuations of energy levels.
For the present double-dot setup, a controllable dephasing mechanism
can be introduced in experiments by performing a ``which-path'' detection
using a nearby quantum point contact (QPC) \cite{SG97,WM97,Buk98}.
The dephasing effect of such a ``which-path" detection
can be accounted for by adding a Lindblad term \cite{SG97}
\begin{equation}\label{LB}
L_{\phi}\rho^{(n)}L^{\dagger}_{\phi}
-\frac{1}{2}L^{\dagger}_{\phi}L_{\phi}\rho^{(n)}
-\frac{1}{2}\rho^{(n)} L^{\dagger}_{\phi}L_{\phi},
\end{equation}
to the r.h.s of \Eq{aa13}. The pure dephasing operator,
$L_{\phi}\equiv\sqrt{\Gamma_{d}}
(|1\rangle_{D}\langle 1|-|2\rangle_{D}\langle 2|)$,
describes the energy fluctuations
of state $|1\rangle_D$ and $|2\rangle_D$ (in the original basis).
Moreover, the dephasing rate can be carried out as \cite{SG97}:
$\Gamma_{d}=(\sqrt{T}-\sqrt{T^{'}})^{2}V_{d}/2\pi$, where $V_{d}$
is the bias voltage across the QPC, and $T$ and $T^{'}$ are the
respective transmission probabilities, depending on which dot
among the two is occupied.
Note also that, since the dephasing does not affect electron's counting,
the above dephasing term is not related to $\rho^{(n-1)}$,
but only to $\rho^{(n)}$.

\subsection{Results and Discussions}

In the following, we restrict our numerical simulation to a symmetric setup,
i.e., $\Gamma_L=\Gamma_R=\Gamma$, and use $\Gamma$ ($\Gamma^{-1}$ )
as the units of energy (time).
For the case of finite counting time ($t_c$), applying the scheme
described in Sec.\ II (C), we identify the LD function
from $\lambda(x,t_c)={\cal F}(x,t_c)/t_c$ and simply denote it as $\lambda(x)$.
Similar simplified notation will be used also in the following for the
$x$-dependent (``$x$-{\it ensemble}") current $I(x)$ and Fano factor $F(x)$.

In Fig.\ 2 we display the LD function $\lambda(x)$
and the associated $x$-ensemble current $I(x)$.
We consider a relatively long counting time $t_c= 150 \Gamma^{-1}$,
and assume the dephasing rate $\Gamma_d=0.01\Gamma$.
Drastically differing from the above result of single dot, here, the LD
function (solid lines in Fig.\ 2) reveals a phase transition behavior,
when crossing through a $\phi$-dependent critical point.

To understand this behavior, we refer to the analytic solution
obtained in Sec.\ IV (B), under the ideal case of $\Gamma_d=0$.
Since the eigenvalues $\lambda_{2,3,4}$ are positive,
the smallest eigenvalue, which has dominant contribution
to the LD function in long counting-time limit, should be either
$\lambda_1$ or $\lambda_5$, depending on the sign of $\lambda_5$.
We can easily check that, as a function of the ``dynamical field" $x$,
$\lambda_5$ can be either {\it positive} or {\it negative},
with a crossover at $x=2\ln|\cos\frac{\phi}{2}|\equiv x_c(\phi)$.
That is, for $x>x_c$, $\lambda_5$ is positive,
while for $x<x_c$ it is negative.
This indicates that the LD function
$\lambda (x)=0$ for $x>x_c$,
and $\lambda (x)=\lambda_5(x)$ for $x<x_c$.

\begin{figure}[h]
\begin{center}
\includegraphics[width=8cm]{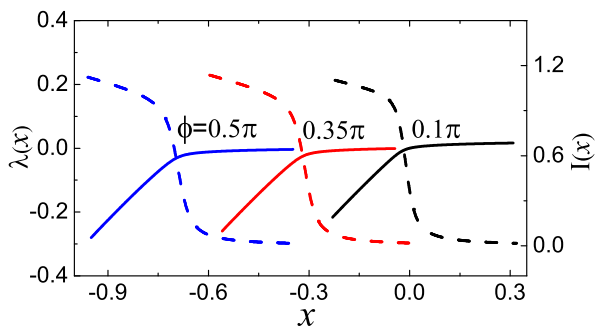}
\caption{\label{fig2}
(color online)
The LD function $\lambda(x)$ (solid lines)
and the $x$-dependent current $I(x)$ (dashed lines)
for a double-dot Aharonov-Bohm interferometer
in strong Coulomb blockade regime.
We assumed a symmetric setup with $\Gamma_{L}=\Gamma_{R}=\Gamma$
and $E_1=E_2$, a relatively long counting time $t_{c}=150\Gamma^{-1}$,
and a dephasing rate $\Gamma_d=0.01\Gamma$.  }
\end{center}
\end{figure}

As a further complementary analysis, let us evaluate the distribution
function $P(n,t)$ in long time limit.
Owing to the assumed Coulomb blockade effect,
we know that once the (transformed) upper dot is occupied,
i.e., the DD is in the state $\tilde{d}^{\dagger}_2|0\ra$,
the current would be completely blocked.
Denoting the tunneling probability from the left electrode
to the upper (lower) dot as $p_{2(1)}$,
the probability with ``$n$" electrons transmitted through
the DD (actually through the lower dot) simply reads
$P(n)=(p_1)^n p_2$, before the transport channel is blocked.
This distribution function defines the entire ensemble
of trajectories, and supports as well a straightforward
``subensemble" LD analysis.
Noting that
$p_{1(2)}=\tilde{\Gamma}_{1(2)L}(\phi)/[\tilde{\Gamma}_{1L}(\phi)
+\tilde{\Gamma}_{2L}(\phi)]$, for the symmetric setup
$\bar {t}_{1 L}=\bar {t}_{2 L}=\bar {t}_{L}$
and $\bar {t}_{1 R}=\bar {t}_{2 R}=\bar t_{R}$,
we obtain a simple result:
$p_1=\cos^2(\frac{\phi}{2})$, and $p_2=\sin^2(\frac{\phi}{2})$.
Now, let us consider the LD transformation,
$P(x)=p_2\sum^{\infty}_{n=0}e^{-xn}(p_1)^n$.
This infinite series has a {\it convergent} sum,
$P(x)=\sin^2(\frac{\phi}{2})[1-e^{-x}\cos^2(\frac{\phi}{2})]^{-1}$,
under the condition $e^{-x}\cos^2(\frac{\phi}{2})<1$.
This convergence condition gives $x>2\ln|\cos(\frac{\phi}{2})|$,
implying that the LD function ($\lambda(x)$) is zero in this region,
being fully consistent with the result based on the eigenvalue analysis.

In Fig.\ 2, we also plot the $x$-dependent current $I(x)$,
by the solid lines. This re-sampled current over a subensemble
of trajectories shows a switching behavior
around the particular point $x_c(\phi)$.
In the ideal case of $E_1=E_2$, $\Gamma_d=0$ and $t_c\rightarrow\infty$,
the switching behavior of $I(x)$ deforms to a discontinuous jump,
owing to the singularity of $\lambda(x)$ at $x_c(\phi)$.
This feature clearly indicates a first-order dynamical phase transition,
revealing a rapid crossover between two distinct
{\it dynamical phases} across $x_c$:
an {\it active} phase given by $x<x_c$ with trajectories having large $n$,
and an {\it inactive} phase by $x>x_c$ with trajectories having small $n$.
At the {\it critical} point $x_c(\phi)$, the two dynamical phases {\it coexist}.

\begin{figure}[h]
\begin{center}
\includegraphics[width=8cm]{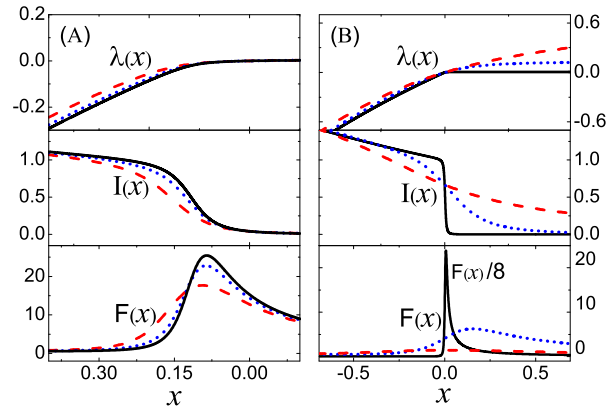}
\caption{(color online)
Effects of the counting time $t_c$ and the dot-level offset $\Delta$.
In (A), for a given $\Delta=0.1\Gamma$ and $\phi=0.25\pi$,
we display the results of $t_c=150\Gamma^{-1}$ (blue dotted lines)
and $50\Gamma^{-1}$ (red dashed lines) against that of
the long $t_c$ limit (black solid lines).
In (B), in the limits of long counting time and $\phi=0$,
we compare the results of $\Delta=0.1\Gamma$ (black solid lines),
$0.5\Gamma$ (blue dotted lines) and $\Gamma$ (red dashed lines).
In both (A) and (B), we assumed $\Gamma_{L}=\Gamma_{R}=\Gamma$
and $\Gamma_d=0$. The latter assumption is mainly for the purpose of
showing the {\it pure} effects of the counting time and the level offset. }
\end{center}
\end{figure}

In Fig.\ 3 (A) and (B) we display, respectively, the effects
of the counting time ($t_c$) and the dot-level offset ($\Delta$).
In Fig.\ 3 (A), as a comparison to the result of long $t_c$ limit
(black solid lines), we plot the LD function, the current
and the Fano factor for two finite counting times,
by the blue dotted and red dashed lines.
We see that, as expected, a shorter counting time will result in
a milder crossover behavior.
As a similar comparison, in Fig.\ 3(B) we show the effect of
the dot-level offset. Since in the transformed basis the two dots
are coupled with a tunneling amplitude proportional to $\Delta$,
we then understand that the dynamical phase
transition will be less drastic as we increase $\Delta$.

\section{Summary}

Starting with a number-conditioned master equation, we formulated
an efficient large-deviation approach for mesoscopic transports.
Formally, it appears that the large-deviation approach is similar to the FCS.
However, a large-deviation analysis is capable of
capturing more complete information: not only for the
{\it typical} trajectories, but also for the {\it rare} ones.
Through an inspection on the continuous change
of subensemble statistics of trajectories,
richer dynamical behaviors can be revealed.
Methodologically, in addition to the eigenvalues analysis
in long time limit, we also presented a convenient
scheme of calculating the large-deviation function
for arbitrary counting times.

As illustrative applications, we considered two examples: one is
the transport through a single dot, another is through a double dots.
Even for the simple system of single dot,
we found that the symmetry of the tunnel coupling
to the leads would strongly affect the statistics of the trajectories,
which is indeed outside the scope of the usual FCS analysis.
Very interestingly, similar to the ``unexpected" finding uncovered
in the quantum optical system of driving two-level atoms \cite{Gar10},
here a symmetric tunnel-coupling in our single-dot transport system
can as well result in a (``time" or ``tunneling-rate")
scale-invariance behavior in the statistics of trajectories.
For the double-dot setup, our large-deviation analysis reveals
a more profound behavior, say, the dynamical phase transition,
which is, essentially, induced by the Coulomb correlation
and quantum interference, and were analyzed in detail
by combining numerical simulations with some analytic solutions.

The large-deviation approach, owing to its ability to
analyze the subensemble statistics of trajectories,
is anticipated to apply to various scenarios,
e.g., from electronic nano-devices to femtochemistry.
In particular, its systematic applications to important
mesoscopic transport systems are extremely promising,
which is seemingly a new area waiting for exploitation.

\vspace{0.5cm}
{\it Acknowledgments.}---
This work was supported by the NNSF of China under grants
No.\ 101202101 \& 10874176, and the Major State Basic Research Project.

%



\end{CJK*}
\end{document}